\documentclass[%
preprint,%
]{revtex4-1}

\usepackage{graphicx}% Include figure files
\usepackage{dcolumn}% Align table columns on decimal point
\usepackage{bm}% bold math
\usepackage[utf8]{inputenc}
\usepackage[T1]{fontenc}
\usepackage{mathptmx}
\usepackage{etoolbox}

\def\@email#1#2{%
 \endgroup
 \patchcmd{\titleblock@produce}
  {\frontmatter@RRAPformat}
  {\frontmatter@RRAPformat{\produce@RRAP{*#1\href{mailto:#2}{#2}}}\frontmatter@RRAPformat}
  {}{}
}%
\begin{document}

%\preprint{AIP/123-QED}

\title[Photoemission characterization of N-polar III-Nitride photocathodes as bright electron beam source for accelerator applications]{Photoemission characterization of N-polar III-Nitride photocathodes as bright electron beam source for accelerator applications}
% Force line breaks with \\
\author{L. Cultrera }
\email{lcultrera@bnl.gov}
\affiliation{Brookhaven National Laboratory, Upton, NY, USA}

\author{E. Rocco}
\affiliation{ 
College of Nanoscale Science and Engineering, SUNY Polytechnic Institute, Albany, NY, USA}
\author{F. Shahedipour-Sandvik }
\affiliation{ 
College of Nanoscale Science and Engineering, SUNY Polytechnic Institute, Albany, NY, USA}
\author{L. D. Bell}
\affiliation{Jet Propulsion Laboratory, California Institute of Technology, Pasadena, CA, USA}%
\author{J. K. Bae}
\affiliation{CLASSE, Cornell University, Ithaca, NY, USA}%
\author{I. V. Bazarov}
\affiliation{CLASSE, Cornell University, Ithaca, NY, USA}%
\author{S. Karkare}
\affiliation{Department of Physics, Arizona State University, AZ, USA}%
\author{P. Saha}
\affiliation{Department of Physics, Arizona State University, AZ, USA}%
\author{A. Arjunan}
\affiliation{Structured Material Industries Inc., Pitascaway, NJ, USA}

\date{\today}% It is always \today, today,
             %  but any date may be explicitly specified

\begin{abstract}
We report on the growth and characterization of a new class of photocathode structures for application as electron sources to produce high brightness electron beams for accelerator applications. The sources are realized using III-Nitride materials and are designed to leverage the strong polarization field characteristic of this class material while grown in their wurtzite crystal structure to produce a negative electron affinity condition without the use of Cs, possibly allowing these materials to be operated in RF gun. A Quantum Efficiency (QE) of about 1x10$^{-3}$ and a Mean Transverse Energy (MTE) of electron of about 100 meV are measured at the operating wavelength of 265 nm. In a vacuum level of 3x10$^{-10}$ Torr the QE does not decrease after more than 24 hours of continuous operation. The lowest MTE, about 50 meV, is measured at 300 nm where the measured QE is 1.5x10$^{-5}$. Surface characterizations reveal possible contribution to the MTE due to the surface morphology calling for more detailed studies. 
\end{abstract}
\maketitle

\section{\label{sec:level1}Introduction}
The search for robust photocathode materials that can provide bright electron beams with high efficiency and minimal transverse energies and that can withstand the harsh environment of the electron guns is an active topic for the community involved in the fundamental and applied research of electron sources for accelerators applications. Progresses in the realization of electron sources with improved performance will have a direct impact on a number of facilities by enabling their realization or allowing the expansion of the scope and the scientific reach of the experiment that can be performed with the electron beams they can provide \cite{musumeci}.
The photocathodes materials that have demonstrated to be capable of being successfully operated in electron guns for accelerators belong to two main categories: metals, with copper \cite{quian}, magnesium \cite{nakajyo} and yttrium \cite{scifo-y} being essentially the only ones that have found a practical application; and semiconductors. The number of semiconductor photocathodes used in electron guns is much larger than the metals and another subdivision can be made in terms of Positive Electron Affinity (PEA) and Negative Electron Affinity (NEA) materials. The difference between PEA and NEA materials lies in the fact that for the former materials the electron affinity level at the interface with the vacuum lies above the conduction band minimum while it is below that for the latter. Belonging to the PEA category are alkali antimonides photocathodes (Cs$_3$Sb, CsK$_2$Sb, Na$_2$KSb etc…) that have emission threshold in the visible range of the spectrum and alkali telluride (Cs$_2$Te, CsKTe etc…) that have the onset of the emission in the UV region of the spectrum \cite{dowell}. 
The NEA condition is not naturally found in any known bulk semiconductor materials but it is usually enabled by the formation of a strong electric dipole at the vacuum surface. The most widespread method to generate such dipole make use of a sub-monolayer of Cs evaporated over the pristine surface of a semiconductor. The strength of such dipole can be enhanced by alternating the exposure of the surface to an oxidizing agent like oxygen or NF$_3$. This method has been successfully applied to activate to NEA the surface of a number of semiconductors belonging to the III-Vs \cite{bazarov-nea}, III-Nitrides \cite{wang} and more recently also to lower the work function of ordered oxide materials \cite{galdi}. 
Due to the extreme sensitivity of the activating layer with respect to contamination, photocathodes based on effective NEA have a limited lifetime when operated as electron source for accelerators and require stringent vacuum conditions (vacuum levels better than 1x10$^{-11}$ Torr) so that their lifetime is sufficiently long to perform experiments. Furthermore, the activation layer can also be damaged by a moderate increase of sample surface temperatures \cite{Iijima} and by energetic chemical species ionized in the residual gas by the primary electrons and back-streaming toward the cathode surface because of the electric field held applied between cathode and anode \cite{sinclair}. 
With all these limitations any practical operation of NEA activated photocathode has been only possible in high voltage DC photoelectron guns. The large volume required to maintain a sufficiently large distance between the negatively biased cathode electrode and the walls of the vacuum vessel allow for a massive pumping and the reach of the required vacuum levels in the XHV range below the 1x10$^{-11}$ Torr level. Unfortunately, the intensity of the accelerating electric fields at the cathode is at least one order of magnitude lower than the typical accelerating gradient possible in Radio Frequency (RF) guns. This prevents a full leverage of photoemission properties of NEA based photocathode like their simultaneous large QE at near the onset of photoemission combined with the low Mean Transverse Energy characteristic of photoelectron that make these photocathodes ideal candidates for the production of intense high brightness beams \cite{filippetto,bazarov-brigth}. 
GaN, and more in general its ternary alloys and structures with Al and In, offer new opportunities to explore for photocathode applications. In their wurtzitic phase III-Nitrides presents uniquely intense internal field due to self-polarization and piezoelectric effects that can be leveraged to engineer photocathode structures that do not require Cs to be activated to achieve effective NEA \cite{marini}.

\section{\label{sec:level2}The N-polar G\lowercase{a}N photocathode structure}
III-Nitrides of the wurtzite structure can be grown in a number of orientations including Ga-polar (${\textit{c}}$–plane along 0001 direction), N-polar ($\overline{\textit{c}}$–plane along 000$\overline{1}$ direction), non-polar (\textit{m}- and \textit{a}-plane respectively along 1100 and 11$\overline{2}$0 directions), and a variety of semi-polar planes. The magnitude of polarization charge is dependent on the polar angle, with the largest out of plane polarization charge for c-plane III-nitrides \cite{wei}. Ga-polar p-type GaN photocathodes have previously been studied both with the use of a Cs-activation layer \cite{shadi} and with novel Cs-free architectures employing a Si delta-doped surface layer \cite{tripathi}. The Si delta-doped layer and thin n+GaN cap required to stabilize the surface creates a narrow depletion region and increases downward surface band bending. In the Ga-polar orientation, the polarization charge is negative at the surface, which compensates positive depletion charge induced by the structure. In the N-polar orientation, the polarization charge at the surface is positive and compounds with the positive depletion charge to increase the downward surface band bending and lower the vacuum level relative to bulk conduction band energy level \cite{marini}. In fact, utilizing the N-polar orientation n-type concentration in the cap layer can be reduced while maintaining a narrow depletion width, which increases the material quality and efficiency of N-polar photocathode devices. 

\begin{figure}
	\centering
	\includegraphics*[width=0.9\columnwidth]{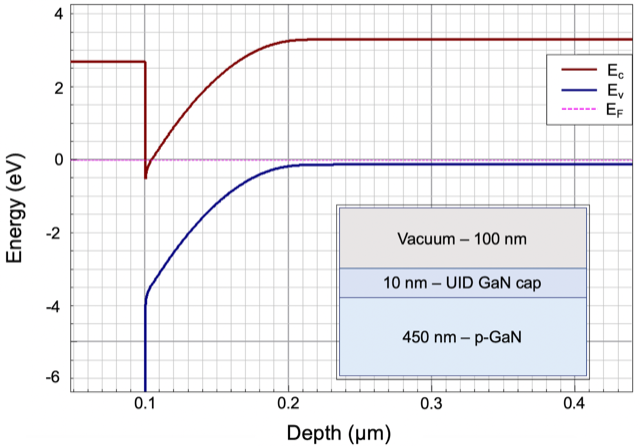}
	\caption{ A simulated energy band diagram of the N-polar photocathode structure utilizing Synposys Sentaurus TCAD software. A schematic of the device structure is shown as an inset. For this simulation an electron concentration of 1x10$^{16}$ cm$^{-3}$ is assumed in the UID GaN cap, and a hole concentration of 3x10$^{17}$ cm$^{-3}$ in the p-GaN layer. The energy band diagram illustrates the sharp downward band bending and narrow depletion width at the surface and capability to achieve NEA through utilizing polarization and band structure engineering.}
	\label{fig0}
\end{figure}

N-polar GaN photocathodes reported here were grown by metal organic chemical vapor deposition (MOCVD). Nominally on-axis 2” sapphire wafers were used as the substrate with 0.2° mis-cut toward the m-plane. An N-polar unintentionally doped (UID) GaN template layer was grown on the sapphire substrate using growth conditions reported elsewhere \cite{marini-2}. The N-polar photocathode structure was then overgrown on the template layer consisting of nominally 450 nm p-GaN with 10 nm UID GaN cap layer\cite{rocco}. A simulated energy band diagram of the device is shown in Fig.~\ref{fig0}  assuming a hole concentration of 3x10$^{17}$ cm$^{-3}$ in the p-type absorbing layer, and an electron concentration of 1x10$^{16}$ cm$^{-3}$ in the UID GaN cap layer. Based on this simulation the structure aims to achieve a narrow depletion width at the surface and NEA. 

Trimethylgallium (TMGa) and ammonia (NH$_3$) were used as the precursors for the p-GaN film and Bis(cyclopentadienyl)magnesium (Cp$_2$Mg) as the dopant source with a V/III ratio of 13,000, TMGa flow of 65 $\mu$mol/min and Cp$_2$Mg flow of 360 nmol/min. The UID GaN cap layer was grown with the same precursors at a V/III ratio of 3,000 and TMGa flow of 80 $\mu$mol/min. Following the completion of the photocathode structure growth the p-type dopants were activated in situ by annealing in nitrogen ambient at 775 °C for 15 minutes. This is completed to remove the hydrogen from the Mg-H complex.

Hexagonal pyramid or ‘hillock’ structures are common on the N-polar surface. The size and areal density of hillock structures can be varied based on the growth conditions \cite{marini-2}, and substrate off-cut \cite{keller}. For the sample studied here, the hillock density varied modestly across the 2” wafer with approximately 1,500 hillocks/cm$^2$ near the wafer flat and 1,000 hillocks/cm$^2$ on the opposite side. Previous studies have shown an improved Mg incorporation efficiency within the sidewalls of the hillock structures leading to improved p-type optical and electrical characteristics for samples with a high density of hillock structures \cite{rocco-2}. Additional details regarding the growth, impurity depth profiles and optical and electrical characteristics of similar structures have been reported by Rocco et al. \cite{rocco,rocco-2}. Two samples were created by cleaving 10x10 mm$^2$ pieces from the 2” wafer, with Sample A taken near the wafer flat with high hillock density and Sample B from the opposite side with lower hillock density. 

\section{\label{sec:level3}Photoemission characterization}
\begin{figure}
	\centering
	\includegraphics*[width=0.9\columnwidth]{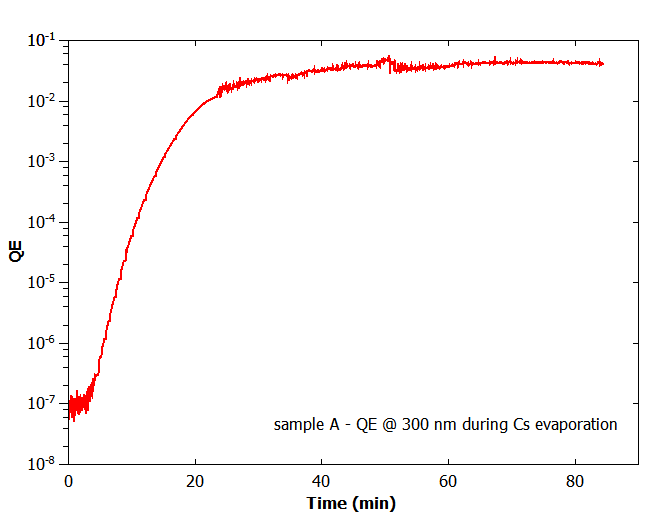}
	\caption{After the initial heat cleaning procedure at 150 °C, the QE of sample A is below the 1x10$^{-7}$ level. Upon exposure to Cs vapors the QE is seen to increase up to approximately 4\%.}
	\label{fig1}
\end{figure}

Prior to photoemission measurements both samples, A and B were etched in a 35\% HCl solution for about 30 s, rinsed with de-ionized water, dried with N$_2$, connected to our sample holder and double bagged in nitrogen atmosphere. Both samples were loaded in the load lock of the main photocathode UHV chamber  within 15 minutes from the initial etch. The load lock was pumped down immediately and after 12 hours 8x10$^{-8}$ Torr vacuum level was reached.
Sample A was moved into the UHV analysis chamber with a base pressure lower that 1x10$^{-10}$ Torr heated for about 12 hours to a temperature of 150 °C and then let it to cool down to room temperature. The initial measurements of quantum efficiency with the surface illuminated with the light produced by a 300 nm LED revealed a very poor QE, below the detection limit of our measurement system which is estimated at 5x10$^{-8}$. The very low QE is possibly indicating that the surface was still contaminated or the threshold energy of the photocathode exceeds the photon energy of 4.1 eV. Hence, an attempt was made to activate the surface with Cs to see if the atoms deposited at the surface could produce a sufficient lowering of the work function so that a measurable photocurrent could be extracted. The Cs atoms were produced using Cs dispenser from SAES Getters. Upon Cs evaporation the photocurrent measured from the sample increased yielding an equivalent QE value of ~4\% at 300 nm (see Fig.~\ref{fig1}). 

\begin{figure}
	\centering
	\includegraphics*[width=0.9\columnwidth]{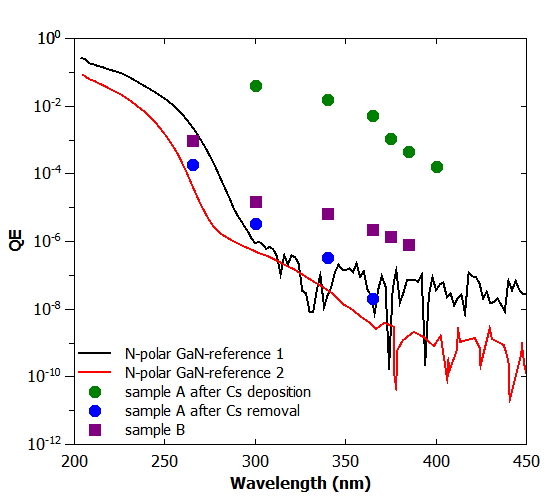}
	\caption{Quantum efficiencies of GaN N-polar structures: sample A was measured after the Cs vapors exposure and after two consecutive heat clean cycles at 600 °C; Sample B was measured after a heat cleaning cycle at 500 °C; spectral responses of similar samples are reported as reference. }
	\label{fig2}
\end{figure}

Measurements of QE were performed using a discrete set of UV LEDs emitting at wavelength of 365, 375, 385, 340, 300 nm and are reported in Fig.~\ref{fig2} (data reported in green dots). 

\begin{figure}
	\centering
	\includegraphics*[width=0.9\columnwidth]{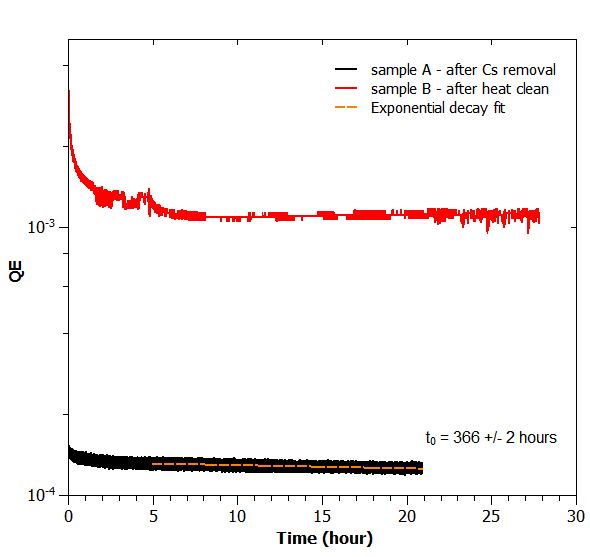}
	\caption{QE as function of the time as measured for sample A and B under illumination with photons at 265 nm. The estimated lifetime, 1/e of QE after the initial transient for sample A is of 366 hours, equivalent to just about 15 days. After the initial transient when QE of sample decreases by a factor 3, the efficiency of sample B is not seen to decrease over 24 hours. Furthermore, the QE at 265 nm exceed the 10$^{-3}$ level.}
	\label{fig3}
\end{figure}

After efficiency measurements were performed, sample A was heated to 600 °C for about 2 hours to remove Cs and other contaminants from the surface. The heating process was repeated twice at 24 hours distance to allow the vacuum to recover, and QE was measured again after the second heating cycle. The second set of measurements on Sample A with Cs-removed, (data reported in blue dots in Fig.~\ref{fig2}) is well aligned with efficiency measurements conducted with a Xe lamp on two samples (N-polar GaN reference-1 and reference-2) grown under similar conditions (continuous black and red lines in Fig.~\ref{fig2}) that were not exposed to Cs vapors. Sample B was moved into the UHV analysis chamber with a base pressure lower that 1x10$^{-10}$ Torr and heated for about 2 hours to a temperature of 500 °C and then let it to cool down to room temperature (for sample A the 1st heat cycle was done at 150 °C for 12 hour). Unlike sample A, sample B was found photoemitting after the heat cleaning and the QE was estimated at the wavelength of 265, 300, 340, 365, 375 and 385 using a set of UV LEDs (data reported as purple dots in Fig.~\ref{fig2}). The obtained values for the QEs are about one order of magnitude larger than the values obtained for sample A but still in the range of efficiency measurements conducted on reference samples, at least at the shortest wavelengths.
The lifetime of samples, defined as the time required for the QE to decay to 1/e of the original value, was estimated by measuring the photocurrent over a span of just over 20 hours. In a base pressure of about 3x10$^{-10}$ Torr the estimated lifetime of Sample A is about 2 weeks with an initial QE at 265 nm just above the 1x10$^{-4}$ level as reported in Fig.~\ref{fig3}. 

The same Fig.~\ref{fig3} reports the measured QE of Sample B during a longer than 24-hour test. After an initial fast decrease (roughly a factor 3 from about 3x10$^{-3}$ to 1x10$^{-3}$) the QE stays essentially stable, if not slowly increasing until the end of the run in a vacuum level of about 3x10$^{-10}$ Torr.

\begin{figure}
	\centering
	\includegraphics*[width=0.9\columnwidth]{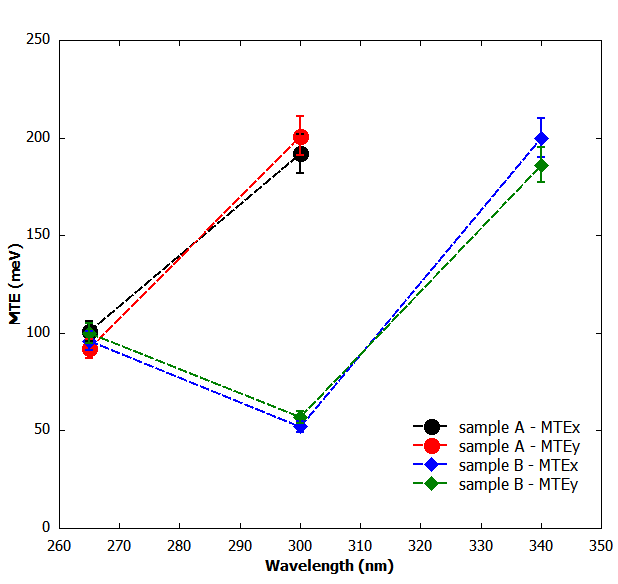}
	\caption{Mean transverse energy of electrons as measured on two perpendicular directions for the two N-polar samples A and B. Because the lower QEs sample A was characterized only at the wavelengths of 265 and 300 nm. Sample B, because of the largest QEs, was also characterized at the wavelength of 340 nm. A non-monotonic trend on the QE is observed for sample B and for bot samples it is observed that the largest MTEs are measured for the longest wavelengths.}
	\label{fig4}
\end{figure}

The MTEs of photoelectrons were measured using the Transverse Energy meter (the TEmeter a low energy photoelectron gun) installed at the photocathode lab of Cornell University using the method of the voltage scan with electron beam energies in the range of 4 to 10 keV \cite{lee}. Due to the low average electron beam current achievable using an LED light source and the relative low QE of the sample at these wavelengths, the MTE measurements for sample A were only possible at the wavelengths of 265 and 300 nm. The results are reported in Fig.~\ref{fig4}. An MTE of about 100 meV is measured at the wavelength of 265 nm and an approximately twice this value is measured at 300 nm. These results are somehow unexpected as usually lower MTEs are associated with lower photon energies. The MTEs of photoelectrons from Sample B were characterized at the three wavelengths of 265, 300 and 340 nm. While similar values were observed at 265 nm with respect to sample A, this set of measurements shows that MTE has a local minimum at the 300 nm wavelength (Fig.~\ref{fig4}). 

\begin{figure*}[htb]
	\centering
	\includegraphics*[width=450pt]{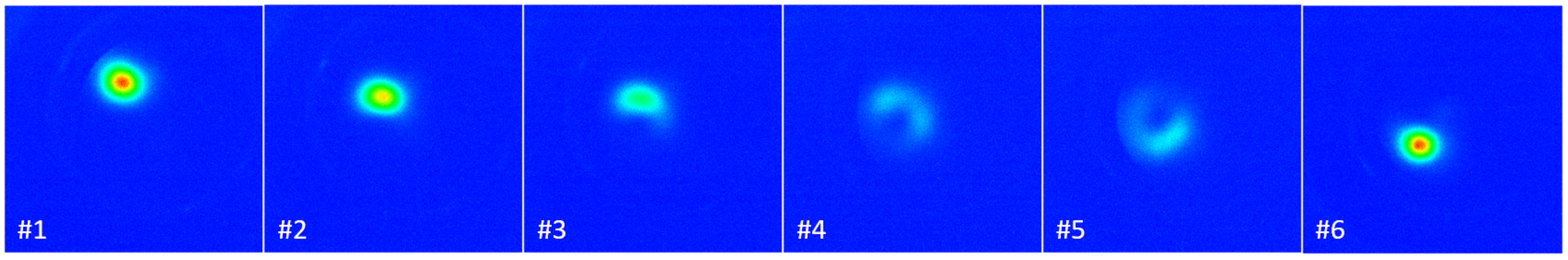}
	\caption{Electron beam intensity profiles observed on the Ce:YAG scintillator screen as the laser beam is vertically scanned on the surface of the sample A in the Transverse Energy meter. The settings of the laser and camera used were kept identical for all the frames.}
	\label{fig5}
\end{figure*}

As the laser beam was scanned over the surface of Sample A during the initial stages of alignment of the instrument it was possible to pinpoint some local areas where the electron beam appeared to have a larger divergence along with reduced intensities. By scanning the laser spot over one of these regions with a laser spot having 25 $\mu$m RMS diameter the sequence of beam images reported in Fig.~\ref{fig5} has been observed. It was not possible to identify, while scanning the laser spot on the photocathode surface of Sample B, a location yielding electron beam images with a hollow center as it was the case for sample A. This seemed to indicate much more uniform properties of photoemission over the accessible area.

\begin{figure}
	\centering
	\includegraphics*[width=0.7\columnwidth]{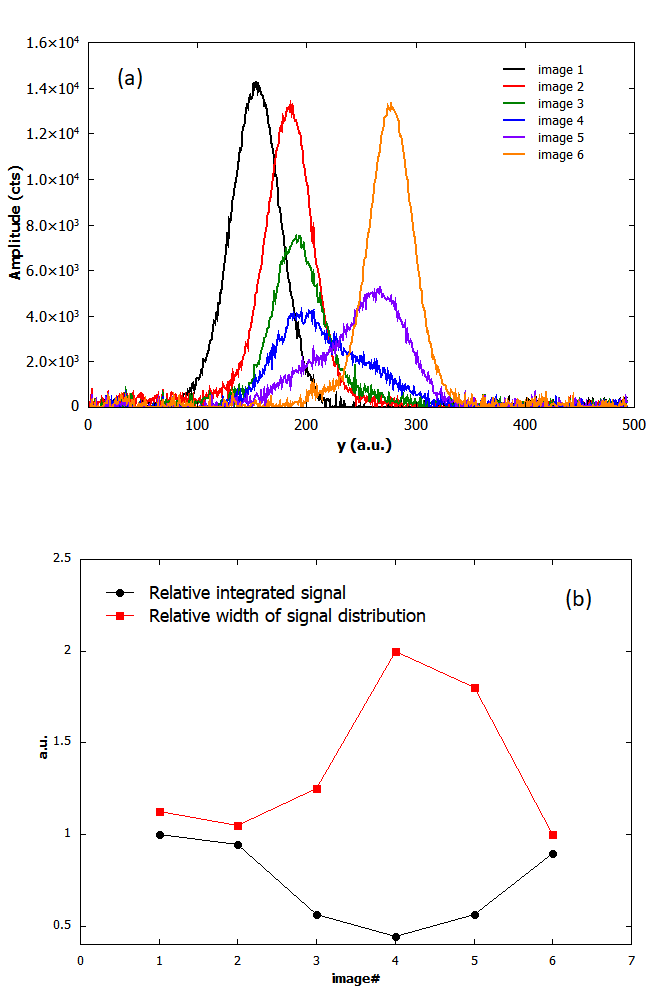}
	\caption{The vertical beam profiles in panel (a) are obtained by integrating the signal along the rows of the images reported in Fig.~\ref{fig5}. The total integrated signal is reported here normalized with respect to the largest signal, which happens to be that of frame 1. Similarly, the width of the signal distribution, obtained by the beam profiles reported in panel (a) are normalized with respect to the smallest one, which happens to be that in frame 6. As the laser spot is moved over the surface area it does encounter an area where simultaneously QE is reduced and the MTE of electron increases as seen in panel (b).}
	\label{fig6}
\end{figure}

The electron beam images in Fig.~\ref{fig5} were recorded using the same settings for the camera. By analyzing the beam profiles obtained by integrating the signal along the vertical direction (see Fig.~\ref{fig6}a) the relative intensities and beam relative width of distribution have been obtained and reported in Fig.~\ref{fig6}b. The data in Fig.~\ref{fig6}b confirm that as the laser beam is moving along the vertical path it encounters a spot where there are both reduced quantum efficiency (as deduced by the decrease of the integrated signal on the scintillator screen) and increased MTE (as deduced by the increase of the width of the signal distribution). Electron beam intensty distributions in frame 4 and 5 of Fig.~\ref{fig5} seems to indicate that a correlation on initial position and momentum of electrons must be present at the photocathode surface so that a hollow electron beam is imaged on the screen once the laser beam spot is centered on the low efficiency emitting area. 

\section{\label{sec:level4}Surface morphology}

\begin{figure*}
	\centering
	\includegraphics*[width=400pt]{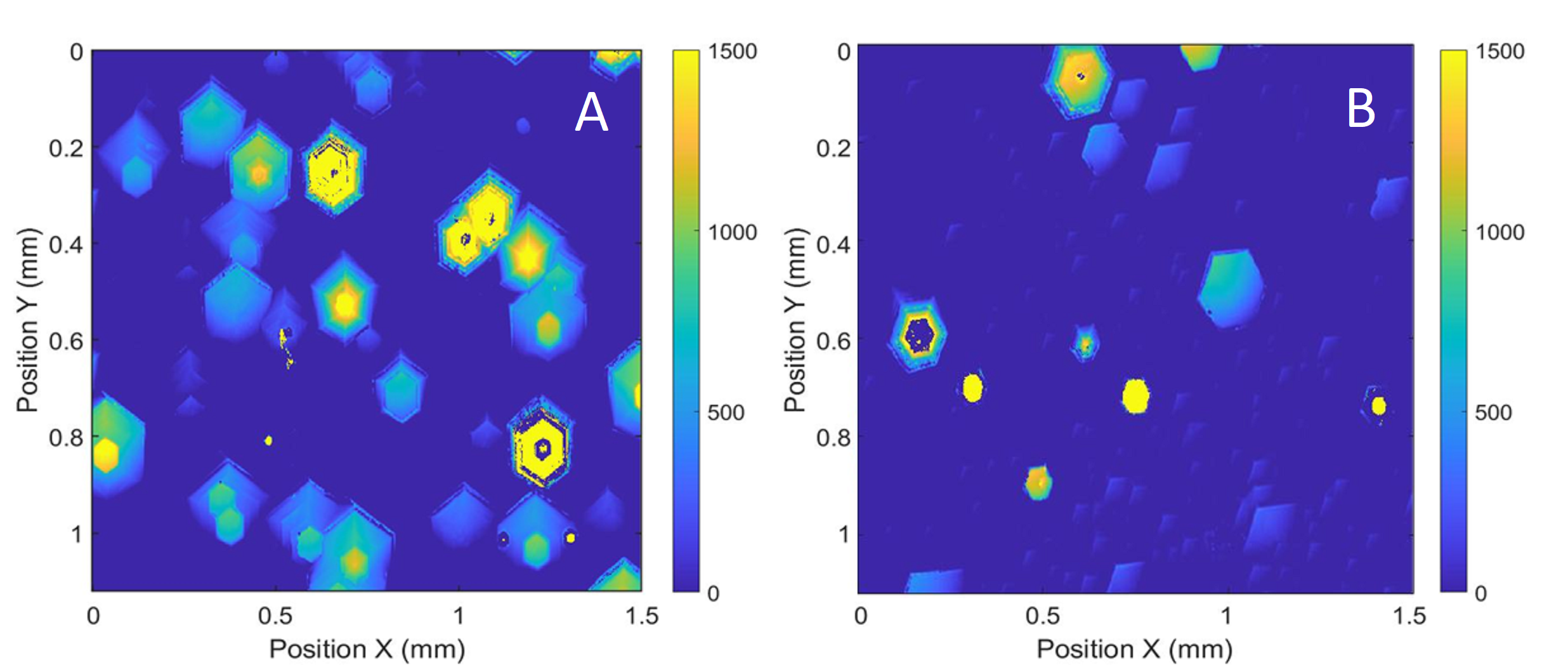}
	\caption{Two dimensional maps of the surface morphology from the two samples A and B. Vertical scale is expressed in nm.}
	\label{fig7}
\end{figure*}

The surface morphology of the samples was investigated using white light interferometry profilometer (Zygo Zescope). Two dimensional maps of representative scanned area belonging to sample A and B are reported in Fig.~\ref{fig7}. Hillock structures, crystallites with a characteristic hexagonal shape, are observed to have developed at the surface of both samples. While the density of hillock structures is within the same order of magnitude on both samples, hillock structures on sample A in general have larger diameters and heights as compared to sample B, and a Sample A has a modestly higher hillock density overall. These structures have been observed to develop during the MOCVD growth of N-polar samples, and  have been associated with increased p-type characteristics by increasing the Mg incorporation efficiency inside the GaN lattice (just above 1x10$^{19}$ cm$^{-3}$) [20]. 

\section{\label{sec:level5}Discussion}

The photoemission properties of the two N-polar photocathode structures are extremely interesting for application as  electron beam source for accelerator purpose. In the case of sample B, the relatively large QEs measured at the largest photon energies, above the 1x10$^{-3}$ level, coupled with the measured MTE of 100 meV and the long lifetime associated with the relative inertness of the surface are surely competitive and outperform many of the metals used in high gradient RF guns. As the photon energy decreases, approaching the threshold of photoemission at 4.1 eV the efficiency lowers in the 10$^{-5}$ level typical of metals like copper but the MTEs measures for the N-polar photocathode a factor 2 smaller than the ones measured for copper cathodes with extremely low roughness \cite{scifo-cu}. At the largest wavelength the rapid increase is on another hand difficult to explain and will require more detailed investigations: a similar behavior was recently observed also for Cs-Te operated at near threshold and attributed to other phases of Cs-Te characterized by a lower emission threshold and lower QE \cite{pierce}. In that study the increase of MTE was observed on photoelectrons extracted with photon energies in the visible range using the light of a supercontinuum source covering a spectral range where QEs were ranging between 1x10$^{-5}$ and 1x10$^{-7}$. Unfortunately, at the time the present photoemission experiments were performed, the lack of a continuously tunable light source in the UV yielding a sufficient photon flux to perform MTE measurements with a fine wavelength scan, only allow us to speculate on the origin of the non-monotonic MTE behavior observed for sample B. 

The spectral response data reported in Fig.~\ref{fig1} for sample B shows that the photoemission is characterized by a low efficiency shoulder located in the range between 300 and 385 nm and by a higher efficiency region at wavelength shorter than 300 nm. At this time, we do not have sufficient information allowing us to pinpoint the origin of photoemission generating low efficiency shoulder to a specific mechanism (e.g. surface states, deep defects level etc.). 
A simplified model allows us to estimate the MTE of photoelectron as 1/3 of the excess energy, which is the difference between the photon energy and the photoemission threshold \cite{bazarov-mte}. This simplified model has been used to describe the photoemission result from Cs-Te near the onset of photoemission reporting an unexpected non-monotonic behavior of the MTE as function of the photon energy \cite{pierce}. These photoemission studies on Cs-Te lead to the conclusion that the non-monotonic behavior of the MTE, which is seen to increase, rather than decrease, as the photon energy falls below a certain threshold was indeed due to the presence of two distinct populations in the electron beam and to an abrupt change of the ratio of them as the photon energy approaches the threshold separating the high and low quantum efficiency spectral regions \cite{pierce}. Similarly to the Cs-Te case, the spectral response of the N-polar GaN seems to be characterized by two regions showing a low efficiency shoulder at longer wavelengths and a high efficiency region at shorter wavelengths (see Fig.\ref{fig1}). Again, the origin of the shoulder at longer wavelengths is not yet unequivocally determined but, if we assume that similarly to the case of Cs-Te we are in presence of two distinct electron populations in the electron beam, when the wavelengths are longer than the threshold for the high QE photoemission region the MTE will progressively increase as they approaches the threshold of the high QE spectral region. Once this threshold is reached and eventually overcome, the fraction of electrons emitted from the low QE shoulder will become less and less important with respect to the total number of electrons and the MTE of the electron beam should initially decrease due to the fact that the largest fraction of photoelectrons has now a smaller excess energy and MTE because of the larger emission threshold of this second population. As the illuminating wavelengths get shorter the MTE will increase again because of the larger excess energy of the electrons.
The formation of a hollow electron beam is observed on the screen when the laser spot illuminates some specific locations and is accompanied by an apparent decreased quantum efficiency and larger momentum spread as deduced from the projected beam profiles reported in Fig.~\ref{fig6}a. The analysis of surface morphology had revealed that the surface of both samples is partially covered with structures with a characteristic diameter comparable with the laser spot size used on our experiments (see Fig.~\ref{fig7}). The structures are 3-D hexagonal based pyramid structures with angled, sloping sidewalls. A detailed study of the effect of these structures on the electron MTEs will require more accurate measurements possibly performed with a smaller laser spot size to correlate the position on the cathode surface with the MTEs and QE. It is expected that such structures will increase the divergence of the beam and hence affect the MTE of electrons \cite{gevorkyan}. The decrease in QE deduced by the reduced integrated intensity on the beam images indicates that the QE performances measured in these samples still have large margins for improvement. Despite the still limited data collected, the overall performance of N-polar GaN photocathodes are already noteworthy and demonstrate that these engineered structures can potentially fulfill the requirement for applications like high repetition rates FELs. Lifetimes in excess of 15 days have been measured indicating the robustness of these advanced structures.
The possibilities for further improvement and new explorations are vast and can leverage decades of R\&D investment on development in the LED industry: emission wavelength tuning can be achieved using ternary alloys with Al and In, efficiency at near threshold can be improved by using resonating Fabry-Perot structures and furthermore by reducing the density of defects using newly available substrates based on single crystal GaN rather than growing on sapphire substrates \cite{kucharski}. 
Future studies of N-polar III-Nitride photocathodes should aim at understanding the origin of the low efficiency shoulder in the photoemission and identifying path for its mitigation so that lower MTEs can be achieved. Growth procedures that can produce smoother surface are likely to be needed as the hillock surface structures might generate increased MTEs and possibly problematic level of field emission currents in very high electric field gradient. Investigation of robustness against known surface contaminants, the possibility of rejuvenating the surfaces only by heating, response time of the photoemission, will also be beneficial to identify possible practical application of this new technology. 

\section{\label{sec:level6}Conclusions}

A new class of photocathode structures, N-polar GaN, has been investigated for the first time as possible candidate to produce bright electron beams. These preliminary results are promising but the range of required characterizations needs to be further expanded to gather more insight from the point of view of both material science and the electron beam production. Simultaneous QE in the range of 10$^{-3}$, with reasonable low MTEs of about 100 meV and long lifetimes already make these materials very interesting. The use of transparent substrates makes them appealing for operation in transmission mode. Furthermore, the concept of N-polar structures can be extended to other III-Nitride structures with reduced threshold for photoemission.

\begin{acknowledgments}
This work was supported by U.S. Department of Energy under contracts DE-SC0012704, DE-SC0021092, DE-SC0020779 and U.S. National Science Foundation under Award PHY-1549132, the Center for Bright Beams.
This research was carried out in part at the Jet Propulsion Laboratory, California Institute of Technology, and was sponsored by the National Aeronautics and Space Administration (80NM0018D0004).
We acknowledge the use of facilities within the Eyring Materials Center at Arizona State University supported in part by NNCI-ECCS-2025490.

\end{acknowledgments}

\section*{Data Availability Statement}
The data that support the findings of this study are available from the corresponding author upon reasonable request.

\bibliographystyle{unsrt}
%\bibliography{aipsamp} % read bib file

%\nocite{*}
%{%
	% "biblatex" is not used, go the "manual" way

%\end{thebibliography}
%}    

\end{document}